\RequirePackage{fix-cm}
\documentclass[smallextended]{svjour3}       
\smartqed  
\usepackage{graphicx}
\usepackage{cite}
\usepackage{amsmath,amssymb,amsfonts}
\usepackage{graphicx}
\usepackage{textcomp}
\usepackage{soul}
\usepackage{url}
\usepackage{cite}
\usepackage{float}
\usepackage[scaled=.92]{helvet}
\newcommand{\R}{\mathbb{R}}

\usepackage{caption}
\usepackage{comment}
\usepackage[ruled,linesnumbered]{algorithm2e}

\usepackage{float}
\usepackage{multicol}
\usepackage{color}
\newcounter{multifig}
\usepackage{thmtools,thm-restate}
\usepackage{multirow}
\usepackage{mathtools, cuted}

\usepackage{lipsum}

\usepackage{array}
\usepackage{makecell}

\usepackage[caption=false, font=footnotesize]{subfig}

\newtheorem{assumption}[theorem]{Assumption}

\def\BibTeX{{\rm B\kern-.05em{\sc i\kern-.025em b}\kern-.08em
		T\kern-.1667em\lower.7ex\hbox{E}\kern-.125emX}}
\begin{document}

\title{Distributed Algorithms for Internet-of-Things-enabled Prosumer Markets: A Control Theoretic Perspective\footnote{\bf{To appear as a chapter in {\em Analytics for the Sharing Economy: Mathematics, Engineering
and Business Perspectives}, Editors: E. Crisostomi et al., Springer, 2019  (forthcoming book).}}\thanks{The work is partly supported by Natural Sciences and Engineering Research Council of Canada grant no. RGPIN-2018-05096, Danish ForskEL programme (now EUDP) through the Energy Collective project (grant no. 2016-1-12530), and by Science Foundation Ireland grant no. 16/IA/4610.}
}

\titlerunning{Distributed Algorithms for IoT-enabled Prosumer Markets}        

\author{Syed Eqbal Alam         \and
        Robert Shorten  \and 
        Fabian Wirth \and
        Jia Yuan Yu
}


\institute{Syed Eqbal Alam \and Jia Yuan Yu \at
              Concordia Institute for Information Systems Engineering,
              Concordia University, 
              Montreal, Quebec, Canada
           \and
           Robert Shorten  \at
             School of Electrical, Electronic
             and Communications Engineering, University College Dublin, Dublin, Ireland
            \and
            Fabian Wirth \at
            Faculty of Computer Science and Mathematics, University
            of Passau, Passau, Germany 
}

\date{Received: date / Accepted: date}

\maketitle
\begin{abstract}
 {\em Internet-of-Things} (IoT) enables the development of sharing economy applications. In many sharing economy scenarios, agents both produce as well as consume a resource; we call them {\em prosumers}. A community of prosumers agrees to sell excess resource to another community in a prosumer market. In this chapter, we propose a control theoretic approach to regulate the number of prosumers in a prosumer community, where each prosumer has a cost function that is coupled through its time-averaged production and consumption of the resource. Furthermore, each prosumer runs its distributed algorithm and takes only binary decisions in a probabilistic way, whether to produce one unit of the resource or not and to consume one unit of the resource or not. In the proposed approach, prosumers do not explicitly exchange information with each other due to privacy reasons, but little exchange of information is required for feedback signals, broadcast by a central agency. In the proposed approach, prosumers achieve the optimal values asymptotically. Furthermore, the proposed approach is suitable to implement in an IoT context with minimal demands on infrastructure. We describe two use cases; community-based car sharing and collaborative energy storage for prosumer markets. We also present simulation results to check the efficacy of the algorithms.
\end{abstract}

\section{Introduction and Setting} 
Recently, consumers across a range of sectors have started to embrace shared ownership of resources and services with guaranteed access, as opposed to more traditional business models that 
focus on sole-ownership only. The reasons for this trend are multi-faceted and range from societal issues, such as the need to reduce wastage, and more general environmental concerns \cite{Narasimhan2018, Hamari2016, Lan2017, Chen2016}, to pure monetary opportunities
arising from increased connectivity (and the ability that this gives to advertise the availability of unused resources and services) \cite{Crisostomi2018}. Well-known examples of successful companies 
building {\em sharing economy} products include Airbnb (hospitality), Lyft (ride sharing) \cite{Fraiberger2015}, Bird and Lime (scooter sharing), Mobike (bike sharing) \cite{Lan2017}, and Google (Google reviews---information sharing).\newline 

Roughly speaking, several types of sharing application classes are discerned (as described in \cite{Crisostomi2018}).\newline 

\begin{itemize}
	\item[A.] {\bf Opportunistic sharing:} Services based on opportunistic sharing of resources exploit large-scale availability of either unused resources or obsolete business models or both. Examples of products in this area include the parking application JustPark (\url{www.justpark.com}) and the peer-to-peer car sharing services Getaround (\url{www.getaround.com}). The key enablers for such products are mechanisms for informing agents of available resources, their delivery, and payments. \newline 
	
	\item[B.] {\bf Federated negotiation and sharing:} Here, groups of agents come together to negotiate better contracts with utilities (electricity, gas, water, health), or to provide mutually beneficial services such as collaborative storage of energy. The key enablers for such products are mechanisms for grouping communities and for enforcing contractual obligations for federations of like-minded consumers. 
	 \newline
	
	\item[C.] {\bf Bespoke sharing:} In this case, products are designed with the specific objective of being shared, rather than for sole ownership. A basic example of such systems is devices and services that allow sharing of a single electric charge-point by several users. Other examples include time-shared apartments or cars that are owned by several people rather than a single person \cite{Crisostomi2018}. \newline
	
	\item[D.] {\bf Hybrid sharing:} Finally, opportunities also exist for sharing economy to support the regular economy. We readily find examples of such systems in the hospitality industry, where ad-hoc sharing economy infrastructure (spare rooms in local houses) can be used as a buffer to accommodate excess demand in the regular economy (hotels).\newline 
\end{itemize}
 The common characteristic in all of the above application classes is the ability for community-wide communication and actuation, both to enable services to be bought and sold, and so that contracts can be enforced. \newline
 
   The development of sharing economy applications \cite{Huckle2016, Kortuem2016} is facilitated by Internet-of-Things (IoT), for example, IoT helps to perform secure payments, to track the location and the condition of an object, to list a few. Interested readers can find several IoT-based applications in \cite{Atzori2010, Fuqaha2015} and the papers cited therein.
 Therefore, while the value of the sharing economy is not in question \cite{Goudin2016, Hamari2016} and while many of the essential infrastructural elements needed for the deployment of such systems are being developed rapidly, there is an additional requirement for structured platforms to enable distributed {\em community-wide} buying and distributed {\em community-wide} selling. Currently, such platforms are at a very early stage of development with significant opportunities for improvement.\newline 

Our objective in this chapter is to address this deficit partially and to develop tools to support the design of community-based prosumer markets. We define {\em prosumers} as agents that both produce and consume a resource \cite{Ritzer2010}. Specifically, we are interested in developing {\em light} algorithms that can easily be deployed on modest {IoT} platforms, and that can be used to support distributed community-wide buying and selling of resources. Here, by {\em light}, we mean algorithms that place low demands on the infrastructure, both in terms of computational power, and actuation and connectivity requirements of individual prosumers. A fundamental requirement is also that such algorithms are scale-free in the sense that they can operate across a range of community sizes; from small communities of a few prosumers to larger communities made up of very many prosumers. These constraints are directly related to the challenges associated with uncertainties that arise in the context of sharing economy problems. Typically, at any time instant, one does not know how many prosumers are participating in the sharing scheme; whether prosumers can or are willing to communicate with each other (perhaps due to privacy considerations), and whether enough computational power is available to the whole network to allocate resources in real-time optimally. A further complication is that we would like any scheme that we develop to be backward compatible with old IoT platforms that support only essential interaction between prosumers and infrastructure. Thus, there is considerable interest in developing {\em light} algorithms that place only modest demands on infrastructure, yet can be used to implement complex policies in the face of the uncertainties mentioned above.\newline 

Given this context, we are particularly interested in situations where communities come together to purchase and sell related commodities simultaneously. Such systems arise, for example, in energy systems where agents (prosumers) both produce and consume energy \cite{Ritzer2010, Moret2018, Agnew2015}. 

\section{Prosumer Markets and Communities}
{\em Prosumers} are the agents that both produce and consume resources \cite{Ritzer2010, Patel2017}. We are interested in prosumer markets that facilitate a community of prosumers for distributed production and consumption. Such markets are emerging rapidly in energy sector \cite{Inderberg2018, Schill2017}, but also in other areas such as shared mobility \cite{Grijalva2011}. Parag and Sovacool \cite{Parag2016} classify prosumer markets according to three network architectures \footnote{In the network architectures, {\em p} represents a prosumer.}. \newline

\begin{itemize}
	\item[(i)] {\em Prosumer-to-prosumer model:} In this peer-to-peer model, prosumers interact (buy or sell resource) directly with each other as depicted in Figure \ref{Diag_PPmodel}. This model is widespread. For example, consider the case of car sharing platform Turo. Here, car owners list their cars on Turo sharing platform, and the riders book the cars of their choice through this platform for a certain period with a fee. A similar, peer-to-peer model is proposed in \cite{Zhang2018} for energy trading in microgrids.
	\newline    
	\begin{figure}[ht]
		\centering
		\includegraphics[width=0.45\linewidth]{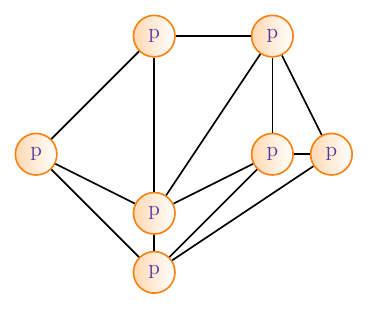}
		\caption{Prosumer-to-prosumer model. Here {\em p} represents a prosumer. Figure adapted from \cite{Parag2016}.}
		\label{Diag_PPmodel}
	\end{figure}
	
	\item[(ii)] {\em Prosumer-to-firm model:} In this model, prosumers interact with a local firm directly. There are two types of prosumer to firm models, {\em prosumer-to-interconnected-firm} model, and {\em prosumer-to-isolated-firm} model.  In the prosumer-to-interconnected-firm model, prosumers are connected to a local firm, which may be connected to the main firm as presented in Figure \ref{Pro_model2}(a). For example, suppose that prosumers produce energy from renewable sources and are connected to a microgrid. A prosumer satisfies its energy needs from the microgrid and the energy it produces. If the prosumer produces more energy than it needs; then, it can return excess energy to the microgrid, this microgrid may be connected to the main grid. Whereas, in the prosumer-to-isolated-firm model, prosumers are connected to the local firm, which works in isolation as depicted in Figure \ref{Pro_model2}(b). Example of the prosumer-to-isolated-firm model is Island microgrid \cite{Ribeiro2011} in which prosumers and microgrid work together to fulfill the energy need of prosumers in the Island. \newline 
	
	\begin{figure*} [ht]
		\centering
		\subfloat[]{%
			\includegraphics[width=0.45\linewidth]{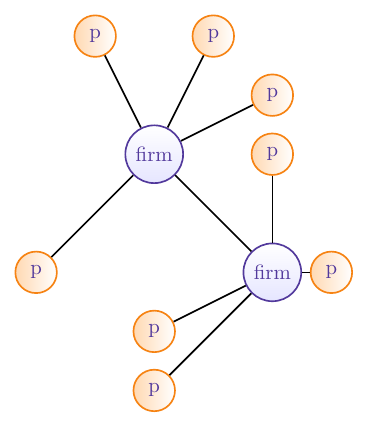}}
		\label{Diag_PF2model}\hfill
		\subfloat[]{%
			\includegraphics[width=0.45\linewidth]{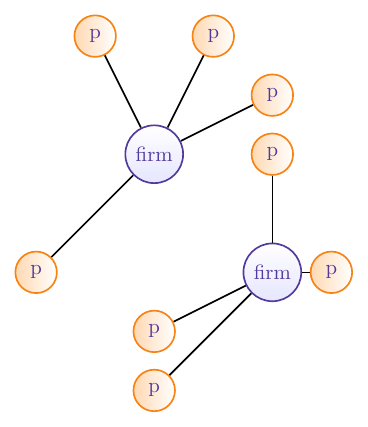}}
		\label{Diag_PF2model_is}\hfill
		
		\caption{ Prosumer-to-firm models---(a) prosumer-to-interconnected-firm model, and (b) prosumer-to-isolated-firm model. Figure adapted from \cite{Parag2016}.}
		\label{Pro_model2} 
	\end{figure*} 
	
	\item[(iii)] {\em Community-based prosumer model:} In this model, prosumers are located in the same geographic location who have similar resource needs and resource production pattern; more generally, they share common goals and interests. These prosumers are grouped to interact with each other and efficiently manage the resource needs of the community, as depicted in Figure \ref{Diag_Commodel}. In this case, communities may also exchange resources with each other. A recent example of community-based trip sharing is found at \cite{Hasan2018} in which the algorithm clusters commuters in communities to optimize car usage. We clarify that, for simplicity, we consider a single prosumer community that interacts with another community ({\em external community}) in the rest of the chapter unless otherwise stated.
	
	\begin{figure}[ht]
		\centering
		\includegraphics[width=0.6\linewidth]{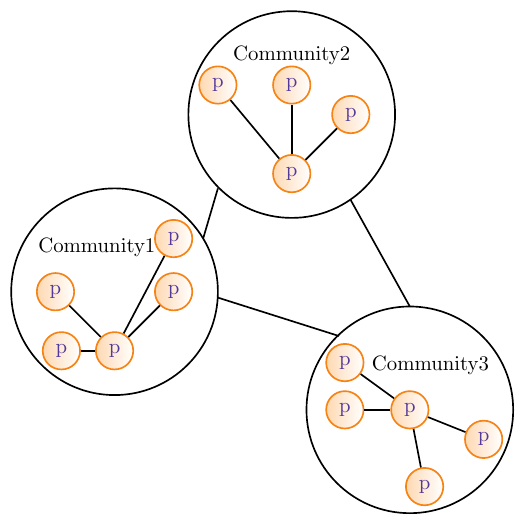}
		\caption{Community-based prosumer model. Figure adapted from \cite{Parag2016}.}
		\label{Diag_Commodel}
	\end{figure}
\end{itemize}

While work on analytics to help design prosumer markets is still in its infancy, somewhat surprisingly few papers have begun to deal with some of the complex market design issues associated with such systems \cite{Moret2018}, \cite{Gkatzikis2014}, \cite{Iosifidis2017}, \cite{Georgiadis2017}. Roughly speaking, these papers deal with two main issues; (i) the existence of market equilibria, and (ii) methods to allocate resources amongst (competing) prosumers. It is in this latter context that this present work is placed. Generally speaking, resource allocation algorithms for prosumer markets have until now been formulated in an optimization context and can be categorized as is traditionally done for classical optimization models. Namely, resources are either allocated centrally \cite{Einav2016} as in the case of Uber, whereby drivers are assigned to the passengers; or in a distributed fashion as in the case of Airbnb, whereby guests and hosts choose each other; or using hybrid of the above two, as in the case of Didi Chuxing \cite{Narasimhan2018}. As we have said, typically, these allocation problems are formulated in an optimization context, paying particular attention to the certain constraints that arise in the sharing economy. These include privacy of the individual, fair allocation of resources, the satisfaction of service level agreements, and ever-increasing regulatory constraints (for example, in the case of Airbnb).\newline

Our contribution in this chapter is to address these problems using a different approach. Namely, we shall consider these problems (in a control theoretic context) as regulation problems with optimality constraints. In particular, we are interested in applications where prosumer communities both buy and sell resources from or to one, or more external entities. Thus, we take the view that such communities have a contract to both buy and sell a pre-specified level of a resource at a given time instant, and the objective of the allocation algorithm is to ensure that these levels of demand are met. Given this basic setting, we ask the question as to whether this can be done without any explicit exchange of information between individual prosumers in situations where prosumers take only binary decisions (buy or sell), and whether given these constraints, optimal use of the resource can be realized. We shall see in the next section that it is indeed possible to formulate the problem and develop distributed algorithms that achieve all of these properties, and which can be implemented in an IoT context with minimal demands on infrastructure. 
\subsection{Prior Work}
Recent analytics work on sharing economy has primarily followed two directions; (i) market design and equilibria, and (ii) optimal allocation of resources. Here, we give a very brief picture of some of the most recent work. In \cite{Georgiadis2017}, Georgiadis et al. propose three types of resource allocation mechanisms; centralized, coalition-based (game-theoretic approach), and peer-to-peer. In \cite{Gkatzikis2014}, Gkatzikis et al. propose a collaborative consumption mechanism to minimize the electricity cost of a community, and in  \cite{Tushar2018}, Tushar et al. developed a game-theoretic model for peer-to-peer energy trading. Furthermore, in \cite{Moret2018}, Moret et al. design a  community-based distributed energy collective market model that helps energy prosumers to optimize their energy resources and to achieve social welfare of the community. More recently, Iosifidis and Tassiulas \cite{Iosifidis2017}, propose optimization techniques for resource exchange and production scheduling for cooperative systems. Courcoubetis and Weber \cite{Courcoubetis2012}, propose mechanisms for the optimal allocation of shared computing resources. In \cite{Hasan2018}, Hasan et al. propose a community-based car sharing model to maximize the trip sharing. To do so, they use mixed-integer programming, graph theory, and clustering techniques. In \cite{Zhang2018}, Zhang et al. propose a game theoretic peer-to-peer energy trading model between local prosumers and distributed energy resources. They show that the model gives rise to an equilibrium between energy production and consumption. In addition to these works, a peer-to-peer equilibrium model for collaborative consumption is proposed by Benjaafar et al. in \cite{Benjaafar2018}. Finally, Grijalva et al. \cite{Grijalva2011}, discuss architecture for prosumer-based distributed control for the electricity grid. We refer interested readers to a recent review paper by Sousa et al., which covers market design, optimization techniques, and interesting future directions at \cite{Sousa2018}.
\subsection{Contribution} We are primarily interested in applications where prosumer communities buy and sell resources via contracts to one or more external entities and must ensure that they meet certain demands in real-time. Thus, the objective of any allocation algorithm is to ensure that these levels of demand are met. Typically, a problem of this nature would be solved in a standard optimization framework. Unfortunately, this approach is not available to us due to the uncertainty that prevails for this system class. For example, the number of prosumers participating at a given time instant may vary, as may the contracted level of prosumption. Also, for reasons of privacy, prosumers may only communicate with each other or with the infrastructure in a limited fashion, making the communication graph unknown {\em a-priori}, from the point of view of algorithm development. Thus, our approach is to formulate these allocation problems in a control theoretic setting, where the effect of these uncertainties and other disturbances can be dealt with using feedback. Our principal contribution is, therefore, to develop distributed control algorithms that can asymptotically achieve optimality. 
\section{Problem Statement} \label{prob_form} 
 Let us assume that a prosumer market consists of a prosumer community and an {\em external community}. The prosumer community has $N \in \mathbb{N}$ prosumers producing and consuming a resource, which agrees to sell its excess resource to the external community. We also assume that the prosumer market has a control unit, which measures the aggregate consumption and production of the resource. It communicates with the prosumer community as well as the external community; we call it a \emph {sharing platform}.
 Additionally, notice that $N$ is not known to the individual prosumers participating in the scheme.  
For simplicity, we assume 
that there is a single resource produced and consumed by all prosumers, though our formulation can easily be extended to multiple resources and multiple prosumer communities.
In our model, the process of production and consumption takes place at discrete time instants $t_0 < t_1 < t_2 < \ldots$, where $t_0=0$. At each time instant, the overall production and consumption of the previous time instant are evaluated and adjusted. Additionally, we assume that communities and external agencies are contracted to, {\em on aggregate}, consume and produce a certain amount of the resource at time instant $t_k$, for $k=0,1,2,\ldots$\newline

We assume that each prosumer has limited actuation and at any time instant $t_k$, for $k=0,1,2,\ldots$, either it consumes one unit of the resource or it does not consume it. Similarly, each prosumer either produces one unit of the resource at a time instant or it does not produce it. Thus, for each prosumer $i$, we denote by $x_i(k)$ the amount of the resource consumed, and by $y_i(k)$ the amount of the resource produced at time instant $t_k$.\footnote{Depending on the application, the processes of production and consumption may be more appropriately modeled on a continuous time-scale. In this case, we interpret time instants $t_k$ at those times in which the prosumption over the interval $(t_{k-1},t_k]$ is accounted for.} \newline

 We also assume that there are constants $C_x\geq0$ and $C_y\geq0$, specifying the
aggregate consumption and production bounds. Notice that the constants $C_x$ and $C_y$ are known to the sharing platform, but not to individual prosumers in the market. 
Thus, at each time instant $t_k$, we require:
\begin{eqnarray}
\sum_{i = 1}^{N} x_i(k) & = & C_x, \label{eq:one} \quad \mbox{and, }\\
\sum_{i = 1}^{N} y_i(k) & = & C_y. \label{eq:two}
\end{eqnarray}

Our primary objective is to ensure that these prosumption bounds are met at each time instant $t_k$, for $k=0,1,2,\ldots$ 
However, we are particularly interested in situations where production and
consumption are coupled together. For example, in communities that are
formed to produce energy, the time-averaged production and consumption of energy might be coupled through battery storage requirements. Furthermore, in the community-based car sharing prosumer market, the average number of delivered and received ``cars" might be coupled through the desired value of utilization of cars; and similarly, production of a resource 
might depend on its consumption. Thus, in these situations, we would like to ensure 
that consumption and production bounds (\ref{eq:one}) and (\ref{eq:two}) are met asymptotically. To formulate this as a long-term requirement, we introduce the time-averaged consumption:
\begin{align} \label{eq:average_eqn_x}
\overline{x}_i(k) \triangleq \frac{1}{k+1} \sum_{\ell=0}^k x_i(\ell), \quad \text{for } i = 1,\ldots, N,
\end{align}
with the time-averaged production $\overline{y}_i(k)$ defined analogously. Now, let $T_i \in \mathbb{R}_+$ be the desired value of utilization of the resource, for $i = 1,\ldots, N$. Thus, depending on the application (refer Section \ref{use_case}, cf. \eqref{eq:desired_util1}), we might require:
\begin{eqnarray}
\label{eq:limit}
\lim_{k \rightarrow \infty} \overline{x}_i(k) + \overline{y}_i(k) & = & T_i, \quad \text{for } i = 1,\ldots, N, 
\end{eqnarray} 
with some additional constraints on $\overline{x}_i(k)$ and $\overline{y}_i(k)$. In several applications, we might require
  for $\alpha_i \in [0,1]$: 
	\begin{align}
	\lim_{k \rightarrow \infty} \overline{x}_i(k) &= \alpha_i  T_i, \quad \text{and}, 
	\lim_{k \rightarrow \infty} \overline{y}_i(k) &= (1-\alpha_i)  T_i, \quad \text{for } i = 1,\ldots, N.
	\end{align}
Fortunately, in problems that we consider, there is flexibility in
satisfying these constraints and it is enough to satisfy that for 
sufficiently large $k$, we have:
\begin{eqnarray}
\overline{x}_i(k)+\overline{y}_i(k) & \approx & T_i,  \quad \mbox{and,}
\end{eqnarray} 
\begin{eqnarray}
 \overline{x}_i(k) \approx \alpha_i  T_i, \quad  \mbox{and,} \quad    \overline{y}_i(k) \approx (1-\alpha_i)  T_i,
\end{eqnarray}
for $i = 1,\ldots, N$.
To formulate this mathematically,
we associate a cost $g_i:(0,1]^2 \to \R_+, (x_i,y_i) \mapsto g_i(x_i,y_i)$ to the deviation of the actual long-term prosumptions. 

\begin{assumption}[Cost function] \label{assump_1}
	The cost function $g_i(\cdot)$ is strictly convex, strictly increasing in each variable, and is continuously differentiable, for all $i$. 
\end{assumption}

Given this basic setting, we are interested in solving the following
optimization problem:

\begin{problem}[Optimization] \label{prob_des}
	\begin{align} 
	\min_{\xi_1, \ldots, \xi_N,\eta_1, \ldots,\eta_N} \quad &\sum_{i=1}^{N} g_i(\xi_i, \eta_i), \label{eq:pd1} \\
	\text{subject to} \quad
	&\sum_{i=1}^{N} \xi_i = C_x, \label{eq:pd2}		\\
	&\sum_{i=1}^{N} \eta_i = C_y,\label{eq:pd3}		\\
	&\xi_i \geq 0, \label{eq:pd4}\\
	&\eta_i \geq 0, \label{eq:pd5}\quad  \text{for }  i=1,2,\ldots,N.
	\end{align}
\end{problem}
We assume that for this optimization problem an optimal point $(\xi^*,
\eta^*) \in \R^{2N}_+$ exists. The optimal point is unique by
Assumption \ref{assump_1} of strict convexity of the cost function $g_i(\cdot)$. As our problem is timed, we aim to design a distributed algorithm determining, for
each time instant $t_k$, values of $x_i(k)$ and $y_i(k)$, such that for the
long-term averages, we have:
\begin{equation}
\label{eq:optlim}
\lim_{k\to\infty} \overline{x}_i(k) = \xi_i^*, \quad  \lim_{k\to\infty} \overline{y}_i(k) = \eta_i^*, \quad \text{for } i=1,\ldots,N.
\end{equation}
Also, it is desirable that the constraints of the optimization
problem are satisfied by $\overline{x}_i(k), \overline{y}_i(k)$, for every $k$, where $\overline{x}_i(k)$
takes the role of the optimization variable $\xi_i$ and $\overline{y}_i(k)$ that of $\eta_i$.\newline

Typically, a problem of this nature can be solved in a standard optimization framework. Unfortunately, this approach is not available to us for many reasons:\newline 
\begin{itemize}
	\item[(i)] The number of prosumers $N$ in the prosumer community, and the constraints $C_x$ and $C_y$ may vary with time, making an offline computation of an optimal solution difficult.\newline 
	
	\item[(ii)] To preserve privacy, we also assume that prosumers
	do not necessarily communicate with each other, and they only
	communicate with the infrastructure in a limited fashion. Thus, even the
	communication graph is unknown {\em a-priori} for this particular
	problem class. In particular, individual cost function $g_i$ is often private and is not shared by prosumers.\newline 
	
	\item[(iii)] We are interested in algorithms that self organize and converge to an optimal solution even in the presence of disturbances in the state information.\newline 
\end{itemize}

Our approach, therefore, is to treat the above problem as a feedback (stochastic) control problem, which changes the formulation in a minor way; namely, we allow:
\begin{eqnarray}
\sum_{i=1}^{N} x_i(k) \approx C_x, \; \mbox{and} \;
\sum_{i=1}^{N} y_i(k) \approx C_y, \quad \text{for all } k.	
\end{eqnarray}	
In other words, we allow the instantaneous prosumption to undershoot or
overshoot the reference values by a small amount. Then, given this
background, and from Assumption \ref{assump_1} for the cost function $g_i$, for all prosumers $i$, we shall demonstrate that an elementary feedback control algorithm can be
devised to solve an approximate version of Problem \ref{prob_des}. As we shall see, this algorithm requires only a few bits of message transfer as intermittent feedback from a control unit (sharing platform) to prosumers in the prosumer market, but no inter-prosumer communication is required. 
\section{Algorithms for Community-based Prosumer Markets} \label{prelim} 
The algorithms proposed in this section are motivated
by the following elementary argument.
We consider only the case in which at time instant $t_k$, either $0$ or $1$ unit of the resource is consumed or
produced, for $k=0,1,2,\ldots$ For the sake of argument, we only
consider the case of pure consumption in this preamble.\newline

Let $z(k)$ denote the number of
times an agent (consumer) consumes a unit resource until time instant $t_k$, where $k=0, 1,
2, \ldots$  Let $\overline{z}(k)\triangleq \frac{1}{k+1} z(k)$, 
denote the time-averaged consumption of the resource until time instant $t_k$. We assume that the consumption at time instant 
$t_k$ is decided by a stochastic procedure, where the probability of consumption of one unit of the resource is a function $p(\overline{z}(k))$. Additionally, we assume that this probability conditioned on $\overline{z}(k)$
is independent of the previous history of the process.
Then:
\begin{eqnarray}
z(k+1) = z(k)+w(k),
\end{eqnarray}  
where $w(k)$ is a random variable taking the value $0$ or $1$. Thus, the following holds true:
\begin{eqnarray}
\overline{z}(k+1) &=& \frac{k+1}{k+2}\overline{z}(k)+\frac{1}{k+2}w(k) \\
& = & \overline{z}(k)+\frac{1}{k+2}\big(w(k)-\overline{z}(k)\big) \label{eq:preamb1}.
\end{eqnarray}
Recall that $p(\overline{z}(k))$ denotes the probability that $w(k)=1$ at time instant $t_k$, for $k=0,1,2,\ldots$
Then, we rewrite \eqref{eq:preamb1} as:
\begin{eqnarray*}
	\overline{z}(k+1) =   \overline{z}(k)+\frac{1}{k+2}\big(p(\overline{z}(k))-\overline{z}(k)\big)+\frac{1}{k+2}\big(w(k)-p(\overline{z}(k))\big).
\end{eqnarray*}
Note that systems of this form are discussed extensively in \cite{Borkar2008}. In particular, the term 
$\frac{1}{k+2}\big(w(k)-p(\overline{z}(k))\big)$ is a martingale
difference sequence and is treated as noise. It is shown in \cite{Borkar2008}
that under mild assumptions, $\overline{z}(k)$ converges almost surely. The basic idea of the remainder of this section is to construct stochastic feedback algorithms that mimic this 
argument. In particular, our basic idea in the sequel is to choose the
probability distribution $p(\overline{z}(k))$, so that the stochastic system 
both solves a regulation problem and also optimization problem of the form of Problem \ref{prob_des}, simultaneously.
\subsection{ Optimality Conditions} \label{opt_cond}

In this subsection, we briefly discuss the optimality conditions for Problem
\ref{prob_des}, using Lagrangian multipliers. These optimality conditions
lead to the state dependent probabilities that we alluded to in the preamble.\newline

Let 
$x=(x_1,x_2,\ldots,x_N)$ and $y=(y_1,y_2,\ldots,y_N)$. Also, let $\mu^1$, $\mu^2$
and $\lambda^1 = (\lambda_1^1,\ldots,\lambda_N^1)$, $\lambda^2 =
(\lambda_1^2,\ldots,\lambda_N^2)$ be the Lagrange multipliers corresponding
to the equality constraints \eqref{eq:pd2}, \eqref{eq:pd3} and the inequality constraints \eqref{eq:pd4}, \eqref{eq:pd5}, respectively.
The Lagrangian of Problem
\ref{prob_des} is defined as $\mathcal{L}: \mathbb{R}^{2N}\times \mathbb{R}^2 \times
\mathbb{R}^{2N} \to \mathbb{R}$, where  
\begin{multline} 
\mathcal{L}({x}, {y}, \mu^1, \mu^2, \lambda^1, \lambda^2)
\nonumber=   \sum_{i=1}^{N} g_i(x_i,
y_i) - \mu^1 (\sum_{i=1}^{N} x_i - C_x) - \mu^2 (\sum_{i=1}^{N} y_i - C_y) \\ + \sum_{i=1}^{N} \lambda_i^1 x_i + \sum_{i=1}^{N} \lambda_i^2 y_i.
\end{multline}
We assume that the optimal value of Problem \ref{prob_des} is obtained
for positive values of consumption and production. We, therefore,
let $x_i^{*}, y_i^{*} \in (0,1]$ denote the optimal point of
Problem \ref{prob_des}. By this assumption, the inequality
constraints are not active, and it follows that the corresponding
optimal Lagrange multipliers are $\lambda^{*1}=0=\lambda^{*2}$.\newline

Additionally, let $\mu^{*1}$, $\mu^{*2}$ be the optimal Lagrange
multipliers for the equality constraints. The first order
optimality condition is that the gradient vanishes, and inspection
shows that the gradient condition decouples. Recall that $\nabla_x g_i(.)$ denotes the partial derivative of $g_i(.)$
with respect to $x_i$ and $\nabla_y g_i(.)$ denotes the partial derivative of $g_i(.)$ with respect $y_i$. Then, we arrive at the following conditions:
\begin{align*}
\nabla_x g_i(x_i^{*}, y_i^{*}) = \mu^{*1}, \quad \text{for } i=1,2,\ldots,N, 
\end{align*}
and,
\begin{align*}
\nabla_y g_i(x_i^{*}, y_i^{*}) = \mu^{*2}, \quad \text{for } i=1,2,\ldots,N. 	
\end{align*}
In other words, we have:
\begin{align}\label{optimality_cond_b1}
\nabla_x g_i(x_i^{*}, y_i^{*}) = \nabla_x g_j(x_j^{*}, y_j^{*}), \text{ for } i,j \in \{1,2,\ldots,N\}.
\end{align} 
The same consensus condition holds for $\nabla_y g_i(x_i^{*},y_i^{*})$, $i=1,\ldots,N$.
We find that the optimal values satisfy all the Karush-Kuhn-Tucker (KKT) conditions, which are necessary and sufficient conditions for optimality of differentiable convex functions (Chap. 5.5.3 \cite{Boyd2004}).
Hence, the derivatives of the cost functions of all prosumers with
respect to consumption as well as production must reach consensus
at the optimal point.\newline

This type of consensus condition has been used in
\cite{Wirth2014, Griggs2016} (single resource case) and \cite{Syed_sm2018} (multi-resource case) to derive place-dependent probabilities that
ensure convergence to the consensus condition and thus, to the optimal
point. 
\subsection{Algorithm for Consumption} Here, we briefly describe the
distributed algorithm proposed in \cite{Griggs2016} for allocating a
single resource to consumers. In this model, no production is taking place, hence the corresponding variables are omitted. Suppose that there are $N$
consumers in a community. We assume that consumer $i$ of the community has a cost function $g_i:(0,1] \to \mathbb{R}_+$, which is strictly
convex, strictly increasing in each variable, and continuously differentiable, for $i =1,2,\ldots,N$. The random variable $x_i(k) \in \{ 0, 1\} $ denotes the consumption of the unit resource for consumer $i$ at time instant $t_k$, for all $i$ and $k$. As before, let
$\overline{x}_i(k)$ be the time-averaged consumption of consumer $i$ until time instant 
$t_k$, for all $i$ and $k$.\newline

The idea is to choose probabilities so as to ensure convergence to the
social optimum and to adjust overall consumption by the community to the reference value (capacity)
$C_x$, by applying a {\em feedback signal} $\Omega(k)$ to the
probabilities. At each time instant $t_k$, the control unit updates $\Omega(k)$
using a gain parameter $\tau >0$, the past aggregate consumption of the resource, and the capacity $C_x$ as described in \eqref{omega_bs1} and then,  broadcasts the new value to all consumers in the community:
\begin{align} \label{omega_bs1}
\begin{split}
\Omega(k+1) \triangleq \Omega(k) -  \tau  \Big (\sum_{i=1}^N x_i(k) -C_x \Big).
\end{split}
\end{align}
After receiving this signal, consumer $i$ responds in a probabilistic way. The probability distribution
$\sigma_i(\cdot)$ is calculated using the time-averaged consumption $\overline{x}_i(k)$ and the derivative $g_i'(\cdot)$ of the cost function $g_i(\cdot)$, as follows:
\begin{align} \label{sigma_bs}
\sigma_i(\Omega(k),\overline{x}_i(k)) \triangleq \Omega(k)
\frac{\overline{x}_i(k)}{
	{g_i'({\overline{x}}_i(k))}}.	
\end{align} 
Notice that the cost function $g_i$ is chosen as increasing function in each variable so that the probability $\sigma_i(\cdot)$ is in the valid range, for all $i$.
Now, consumer $i$ updates its resource consumption at each time instant $t_k$, either by consuming one unit of the resource or not consuming it, as follows:
\begin{align*}
x_i(k+1) =
\begin{cases} 
1 \quad \text{with probability }  \sigma_i(\Omega(k),\overline{x}_i(k));\\ 
0 \quad \text{with probability }  1 - \sigma_i(\Omega(k),\overline{x}_i(k)).
\end{cases}
\end{align*}
Empirical results show that the time-averaged consumption $\overline{x}_i(k)$ converges to the optimal value $x_i^*$ asymptotically.  
\begin{remark}[Integral control action] \label{rem1}
	Equation \eqref{omega_bs1} defines what is called an integral control
	action in tracking control. The overall consumption by the community should ``track'' the
	available capacity, i.e., approximate it. In other words, the objective of the integrator is to ensure
	that the tracking error $e(k) = \sum_{i=1}^N x_i(k) -C_x \approx 0$ asymptotically.  
\end{remark}

\begin{remark}[Consensus of derivatives] \label{rem2}
	The policy for $\sigma_i(\Omega(k),\overline{x}_i(k))$ in \eqref{sigma_bs} is to ensure that asymptotically, $g_i'({\overline{x}}_i(k)) = g_j'({\overline{x}}_j(k))$, for all consumers $i$ and $j$.
	As is discussed in \cite{Griggs2016}, the convergence of the above algorithm and strict convexity of all cost functions (and of course assuming the existence of a feasible solution in the constraint set) is enough to imply that Problem \ref{prob_des} is solved asymptotically for the consumption case. 
\end{remark}
\subsection{Algorithm for Coupled Prosumption} The case of {\em coupled
	prosumption} of a single resource is characterized by the constraint
\eqref{eq:limit}, that is, consumption and production are coupled through the desired value of utilization of the resource. The algorithm follows the case of exclusive consumption
closely taking into account the cost function as discussed in
the problem formulation (Section \ref{prob_form}). Before proceeding, note
that the following discussion extends to an arbitrary number of resources, but
is presented for a single resource, both to aid exposition and to be
consistent with the application class that is the principal consideration
in this chapter. Interested readers can look at \cite{Syed2018} for preliminary results on multi-resource allocation. With this background in mind, following the discussion for
consumption,  let us assume that in a prosumer market there is a community of $N$ prosumers. The prosumer community sells the excess resource to an external community. Furthermore, let $\tau_x$, $\tau_y$ denote the gain parameters
for consumption and production, $\Omega_x(k)$,
$\Omega_y(k)$ denote the {\em feedback signals} for both processes, and
$C_x, \; C_y$ represent the respective contract (capacity)
constraints. We assume the existence of a centralized
control unit (sharing platform) in the prosumer market that can measure the aggregate response of the prosumer community and broadcast a feedback signal in the prosumer market at each time instant $t_k$, for
both consumption and production type. Specifically, the sharing platform
updates the feedback
signal $\Omega_x(k)$, as follows:
\begin{align} \label{omega_b1}
\begin{split}
\Omega_x(k+1) \triangleq \Omega_x(k) -  \tau_x  \Big (\sum_{i=1}^N x_i(k) -C_x \Big ), \quad \mbox{for all } k,
\end{split}
\end{align}
with $\Omega_y(k)$ updated analogously. After receiving a feedback signal
prosumer $i$'s algorithm responds in a probabilistic manner. The
probability that prosumer $i$ responds to the feedback signal is given
by:
\begin{align} \label{prob_x2}
\sigma_{i,x}(k) \triangleq  \Omega_x(k) \frac{\nabla_x{g_i \big (\overline{x}_i(k),\overline{y}_i(k) \big)}}{ \overline{x}_i(k)}, \quad \mbox{for all } i \mbox{ and } k.
\end{align}	
Here, $\sigma_{i,x}(k)$ denotes the probability of prosumer $i$, responding to a demand for consumption of the resource, at time instant $t_k$; similarly, $\sigma_{i,y}(k)$ denotes the probability of prosumer $i$, responding for production of the resource at time instant $t_k$, for all $k$. 
 As in the consumption case, the cost function $g_i$ is chosen as increasing function in each variable, to keep the probabilities $\sigma_{i,x}(k)$ and $\sigma_{i,y}(k)$ in the valid range. However, the definition of $\sigma_{i,x}(k)$ and $\sigma_{i,y}(k)$ is slightly different from the consumption case, described previously. 
Now, prosumer $i$ updates its consumption and production of the resource at each time instant in the following way:
\begin{align*}
x_i(k+1) &= 
\begin{cases}
1 \quad \text{ with probability } \sigma_{i,x}(k);\\
0 \quad \text{ with probability } 1-\sigma_{i,x}(k), \quad \mbox{and,} 
\end{cases}
\\
y_i(k+1) &= 
\begin{cases}
1 \quad \text{ with probability } \sigma_{i,y}(k);\\
0 \quad \text{ with probability } 1-\sigma_{i,y}(k),
\end{cases}
\end{align*}
 for all $i$ and  $k$. Empirically, we observe that the time-averaged consumption $\overline{x}_i(k)$ converges to the optimal consumption value $x_i^*$, and similarly, the time-averaged production $\overline{y}_i(k)$ converges to the optimal production value $y_i^*$ asymptotically, for all prosumers in the community. We describe the algorithm of the sharing platform in Algorithm \ref{Algo:SP} and the algorithm of prosumers of the community in Algorithm \ref{Algo:pros}.
We make the following remarks.		
\begin{remark}[Communication overhead and privacy] \label{rem3}
	There is no explicit communication between prosumers. Thus, the algorithm is low cost in terms of communication and is private.
\end{remark}

\begin{remark}[Probability bounds] \label{rem31}
	The gain parameters $\tau_x, \tau_y$ are small constants chosen to ensure $\sigma_{i,x}(k)$ and $\sigma_{i,y}(k)$ are the probabilities; namely, are in [0,1], for all $i$ and $k$. 
\end{remark}
\begin{remark}[Consensus of partial derivatives] \label{rem32} The well-posedness of our algorithm follows from the assumption of strict convexity of $g_i(\cdot)$, and that the constraint sets are closed and bounded; namely, that there exists unique optimal solutions to Problem \ref{prob_des}. To show that the long-term average values converge to the optimal values, we use the consensus of (partial) derivatives of the cost functions of prosumers as described previously. That is: 
	\begin{align*}
	\lim_{k \to \infty} \nabla_x g_i(\overline{x}_i(k),\overline{y}_i(k)) = \lim_{k \to \infty} \nabla_x g_j(\overline{x}_j(k),\overline{y}_j(k)),
	\end{align*}  	
	and similarly,
	\begin{align*}
	\lim_{k \to \infty} \nabla_y g_i(\overline{x}_i(k),\overline{y}_i(k)) = \lim_{k \to \infty} \nabla_y g_j(\overline{x}_j(k),\overline{y}_j(k)), 
	\end{align*}
	for all $i,j \in \{1,2,\ldots,N \}$.
\end{remark}

\begin{algorithm}  \SetAlgoLined Input:
	$C_x, C_y$, $\tau_x,\tau_y$, $ x_i(k),y_i(k)$, for $k = 0,1,2,\ldots$ and $i = 1, 2, \ldots, N$.
	
	Output:
	$\Omega_x(k+1), \Omega_y(k+1)$, for $k = 0,1,2,\ldots$
	
	Initialization: $\Omega_x(0) \leftarrow 0.06$ and $\Omega_y(0) \leftarrow 0.06^\text{1}$,
	
	\ForEach{$k = 0,1,2,\ldots$}{
		calculate $\Omega_x(k+1)$ and $\Omega_y(k+1)$ as follows and broadcast them in the prosumer community;	
		\begin{align*} 
		\begin{split}
		\Omega_x(k+1) \leftarrow \Omega_x(k) -  \tau_x  \Big (\sum_{i=1}^N x_i(k) - C_x \Big ), \quad \mbox{and,}
		\end{split} \\
		%
		\begin{split}
		\Omega_y(k+1) \leftarrow \Omega_y(k) -  \tau_y  \Big (\sum_{i=1}^N y_i(k) - C_y \Big ).
		\end{split}
		\end{align*}
	}
	\caption{Algorithm of control unit (sharing platform).}
	\label{Algo:SP}
\end{algorithm}
\footnotetext[1]{We initialize $\Omega_x(0)$ and $\Omega_y(0)$ with positive real numbers.}
\begin{algorithm}  \SetAlgoLined Input:
	$\Omega_x(k), \Omega_y(k)$, for $k = 0, 1, 2, \ldots$
	
	Output: $\overline{x}_i(k+1), \overline{y}_i(k+1)$, for $k = 0, 1, 2, \ldots$
	
	Initialization: $x_i(0), y_i(0) \leftarrow 1$ and
	$\overline{x}_i(0) \leftarrow x_i(0)$ and $\overline{y}_i(0) \leftarrow y_i(0)$.  
	
	\While{prosumer $i$ is active at $k = 0, 1, 2, \ldots$}{
		\begin{align*}
		\sigma_{i,x}(k) &\leftarrow \Omega_x(k)
		\frac{ \nabla_x
			g_i \big(\overline{x}_i(k),\overline{y}_i(k) \big)}{\overline{x}_i(k)};\\
		\sigma_{i,y}(k) &\leftarrow \Omega_y(k)
		\frac{ \nabla_y
			g_i \big(\overline{x}_i(k),\overline{y}_i(k) \big)}{\overline{y}_i(k)};
		\end{align*}
		calculate outcome of the random variables;
		\begin{align*}
		x_i(k+1) \leftarrow 
		&\begin{cases}
		1 \quad \text{ w. p. } \sigma_{i,x}(k)\\
		0  \quad \text{ w. p. } 1-\sigma_{i,x}(k);
		\end{cases} \\
		%
		y_i(k+1) \leftarrow 
		&\begin{cases}
		1 \quad \text{ w. p. } \sigma_{i,y}(k)\\
		0 \quad \text{ w. p. } 1-\sigma_{i,y}(k);
		\end{cases}
		\end{align*}
		update $\overline{x}_i(k+1)$ and $\overline{y}_i(k+1)$ as follows;
		\begin{align*}
		\overline{x}_i(k+1) \leftarrow & \frac{k+1}{k+2}
		\overline{x}_i(k) + \frac{1}{k+2} x_i(k+1);\\
		\overline{y}_i(k+1) \leftarrow & \frac{k+1}{k+2}
		\overline{y}_i(k) + \frac{1}{k+2} y_i(k+1);
		\end{align*}		
		
	}
	\caption{Algorithm of prosumer $i$.}
	\label{Algo:pros}
\end{algorithm}
\section{Use Cases} \label{use_case}
We now describe two use cases for community-based prosumer market. The first is a transportation example and concerns a car sharing prosumer market. The second example is from the energy sector and considers a  prosumer market, where produced and consumed energy is coupled through storage constraints.
\subsection{Community-based Car Sharing} \label{car-sharing}
In this use case, we consider a community of $N$ households (prosumers) with several cars, not all of which are required each day. We assume that cars are pooled and shared amongst community members to multiplex and monetize the excess capacity, but that the average aggregate daily community demand for cars is known. We let $T_i$ denote the average number of cars desired to be used by each household $i$, for all $i$. Suppose that $C_x$ cars are required within the community each day, and an excess of $C_y$ cars are made available to an external community each day. For simplicity, we assume that $C_x$ and $C_y$ are fixed. Notice that a household is a prosumer in the sense that it requires cars for its transportation needs and supplies excess car-days to an external community. For each prosumer $i$, let $x_i(k) \in \{0,1\}$ denote that prosumer $i$ requires a car at day $k$ or not, and let $y_i(k) \in \{0,1\}$ denote that prosumer $i$ supplies a car to an external community at day $k$ or not. Thus, we assume that aggregated over the entire community, the demand for {\em shared cars} on a given day $k$ is:
\begin{eqnarray}
\sum_{i=1}^N x_i(k) = C_x,
\end{eqnarray}
leaving the {\em excess} capacity so that $C_y$ cars can be supplied to an external community:
\begin{eqnarray}
\sum_{i=1}^N y_i(k) = C_y.
\end{eqnarray}

Thus, $\sum_{\ell=0}^{k}x_i(\ell)$ is the number of days a car was required by prosumer $i$ over $k$ days, and $\sum_{\ell=0}^{k}y_i(\ell)$ is the number of days that the same prosumer made a car available to an external community. Thus, $\sum_{\ell=0}^{k} \big(x_i(\ell)+y_i(\ell)\big)$ is the total number of days that a car was used as a result of prosumer $i$. Over some days,
prosumer $i$ will require that the time-averaged of this number be equal to the desired value of utilization of cars, $T_i$. For example, if a prosumer has two cars, thus, over seven days period (a week) it has $14$ car-days. Now, suppose that the prosumer only needs access to $10$ car-days over a week. Then, this prosumer might choose $\sum_{\ell=0}^{6} x_i(\ell)$ and $\sum_{\ell=0}^{6} y_i(\ell)$, such that $\sum_{\ell=0}^{6} \big(x_i(\ell)+y_i(\ell)\big)=12$. In this case, the prosumer sells access to two excess car-days over the week (two car-days rather than the maximum of four to provide some 
margin in case that a car is required for personal use more than expected). Let $\overline{x}_i(k)$ denote the average number of days prosumer $i$ requires a car over $k$ days, and let $\overline{y}_i(k)$ denote the average number of days prosumer $i$ makes a car available to an external community over $k$ days, for all $i$ and $k$. Thus, over a long period, this constraint can be scaled by the number of days to yield:    
\begin{eqnarray} \label{eq:desired_util1}
\overline{x}_i(k) + \overline{y}_i(k) \approx T_i,
\end{eqnarray}
with the cost of not achieving this goal captured by a penalty function $g_i(\overline{x}_i + \overline{y}_i - T_i)$. For sufficiently large $k$, we might also require that  $\overline{x}_i(k) \approx \alpha_i T_i$ and $\overline{y}_i(k) \approx (1-\alpha_i) T_i$ with $\alpha_i \in [0,1]$, for all $i$. This latter constraint can be formulated in terms of a cost via a penalty function. For example, in residential areas in Ireland; two car households are common, meaning that households can in principle both consume and produce cars simultaneously, hence act as prosumers. However, this may not always be possible, and prosumers may be required to make alternative arrangements or pay penalties, should they not be able to meet contractual demands. To formulate the cost function,
we associate costs $h_i:(0,1] \to \R_+, x_i \mapsto h_i(x_i)$ to the deviation from $\alpha_i T_i$
and $l_i:(0,1] \to \R_+, y_i \mapsto l_i(y_i)$ to the deviation from $(1-\alpha_i) T_i$, for all $i$. Then, the aim of the sharing platform is to minimize:
\begin{align}
\sum_{i=1}^{N} \big( g_i(\overline{x}_i+ \overline{y}_i- T_i)+h_i(\overline{x}_i - \alpha_i T_i)+l_i(\overline{y}_i - (1-\alpha_i) T_i) \big),
\end{align} 
subject to the additional constraints listed in Problem \ref{prob_des}.
\subsection{Collaborative Energy Storage}
In this use case, we again assume that there are $N$ households in a community participating in a prosumer market, with each household connected to the grid, and also has installed solar panels. 
Every household has batteries to store the energy either from the solar panel or the grid. The households act as prosumers in the sense that they can consume  stored energy as well as sell excess energy for monetary benefits.
Let $x_i(k) \in \{0,1\}$ denote that household $i$ consumes stored energy at day $k$ or not, and let $y_i(k) \in \{0,1\}$ denote that household $i$ sells stored energy to an external community at day $k$ or not, for all $i$. Let $\overline{x}_i(k)$ denote the time-averaged amount of stored energy consumed by household $i$ over $k$ days, and let $\overline{y}_i(k)$ denote the time-averaged amount of stored energy sold by household $i$ to an external community over $k$ days, for all $i$. 
Now, we assume that $\sum_{i=1}^N x_i(k) = C_x,$ be the aggregated consumption of stored energy over the entire community on a given day $k$;
whereas, the excess energy $C_y$ is sold to an external community, with
$\sum_{i=1}^N y_i(k) = C_y$. Furthermore, we assume that each household may have a constraint on the amount of energy stored
in order to realize the above strategy. As in the previous use case, this {\em soft} constraint is captured as follows. Let $T_i$ be the desired amount of energy stored by household $i$. Then, on average, over a sufficiently large given period $k$, household $i$ may expect to store the following desired amount of energy temporarily:
\begin{eqnarray} \label{eq:desired_util2}
\overline{x}_i(k) + \overline{y}_i(k) \approx T_i,
\end{eqnarray}
with the deviation from this goal captured by a penalty function $g_i(\overline{x}_i+ \overline{y}_i - T_i)$. We might also require that roughly speaking, a certain amount of storage is reserved for consumption and a certain amount to sell (production). 
Hence, again, the objective of the sharing platform is to minimize:
\begin{align*}
\sum_{i=1}^{N} \big( g_i(\overline{x}_i+ \overline{y}_i - T_i) + h_i(\overline{x}_i - \alpha_i T_i)+l_i(\overline{y}_i - (1-\alpha_i) T_i) \big),
\end{align*}
subject to the constraints listed in Problem \ref{prob_des}. Notice that the definition of $h_i(\cdot)$ and $l_i(\cdot)$ is same as the previous use case. 
\section{Simulation Results}
In this section, we present simulation results for $99$ prosumers participating in a prosumer market. We assume that these $N = 99$ in the prosumer market are grouped into two prosumer communities. Prosumers $1$ to $50$ are grouped in Community $1$, and Prosumers $51$ to $99$ are grouped in Community $2$. Additionally, each community has a cost function type, and prosumers of a particular community use the cost function type of that community, but with randomized parameter values. Recall that prosumers can produce the resource as well as consume it. Let time instants $t_0<t_1<t_2<\ldots$, represent days, and let the capacity constraints be $C_x = 90$ and $C_y=80$.
Furthermore, let the cost factors $a_i \in \{1,2,\ldots,10\}$ and $b_i \in \{1,2,\ldots,15\}$ be drawn from uniformly distributed random variables. Additionally, let $\delta =11.75$ and $\gamma = 11.65$. Notice that the values of $\delta$ and $\gamma$ are chosen in such a way that the cost function $g_i(\cdot)$ is increasing in each variable. Recall that this is done to keep the probabilities $\sigma_{i,x}$ and $\sigma_{i,y}$ in $[0,1]$. The cost functions are presented as follows: 
\begin{align} \label{cost_fn1}
g_{i}(\overline{x}_i(k), \overline{y}_i(k)) =
\begin{cases}
			&\delta \big( \overline{x}_i(k) + \overline{y}_i(k)\big)+ \frac{1}{4} a_i\big( \overline{x}_i(k) + \overline{y}_i(k) - T_1 \big)^2 + \nonumber \\ &\frac{1}{2} \big(  \overline{x}_i(k) - \frac{1}{2}T_1 \big)^2 + \frac{1}{2} \big( \overline{y}_i(k) - \frac{1}{2}T_1 \big)^2 \hspace{0.3in} \mbox{Community $1$},
	\vspace{0.1in} \\
& \gamma \big( \overline{x}_i(k) + \overline{y}_i(k) \big) + \frac{1}{8} a_i\big( \overline{x}_i(k) + \overline{y}_i(k) - T_2 \big)^2 + \nonumber \\ &\frac{5}{4} b_i\big( \overline{x}_i(k) + \overline{y}_i(k) - T_2 \big)^4 + \frac{1}{2} \big( \overline{x}_i(k) - \frac{1}{2}T_2 \big)^2 + \nonumber \\ &\frac{1}{2} \big(  \overline{y}_i(k) - \frac{1}{2}T_2 \big)^2 \hspace{1.45in} \mbox{Community $2$}.
	\end{cases}
	\end{align}
%
\begin{figure}[h] 
	\centering
	\subfloat[]{%
		\includegraphics[width=0.485\linewidth]{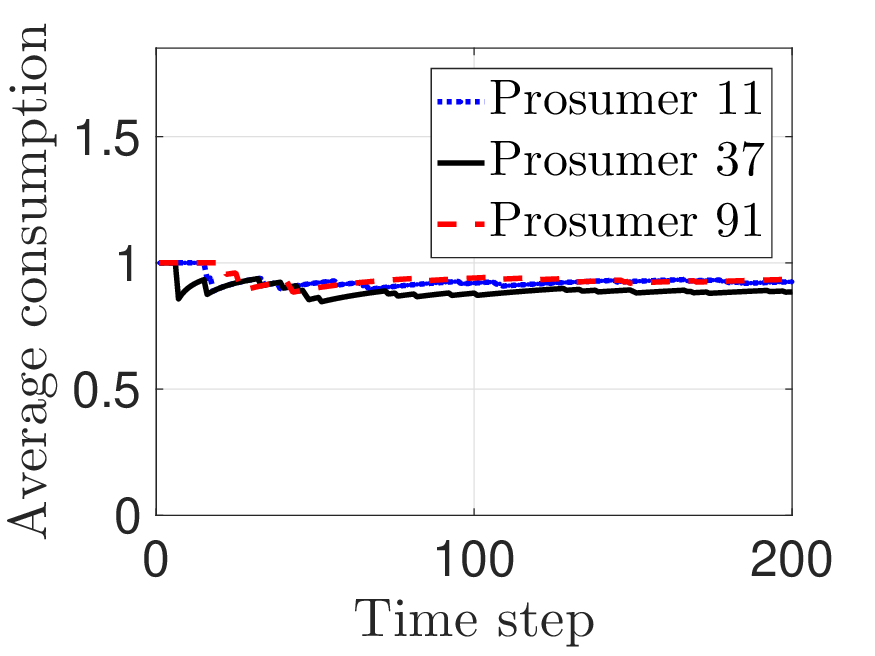}}
	\label{avgx}\hfill
	\subfloat[]{%
		\includegraphics[width=0.485\linewidth]{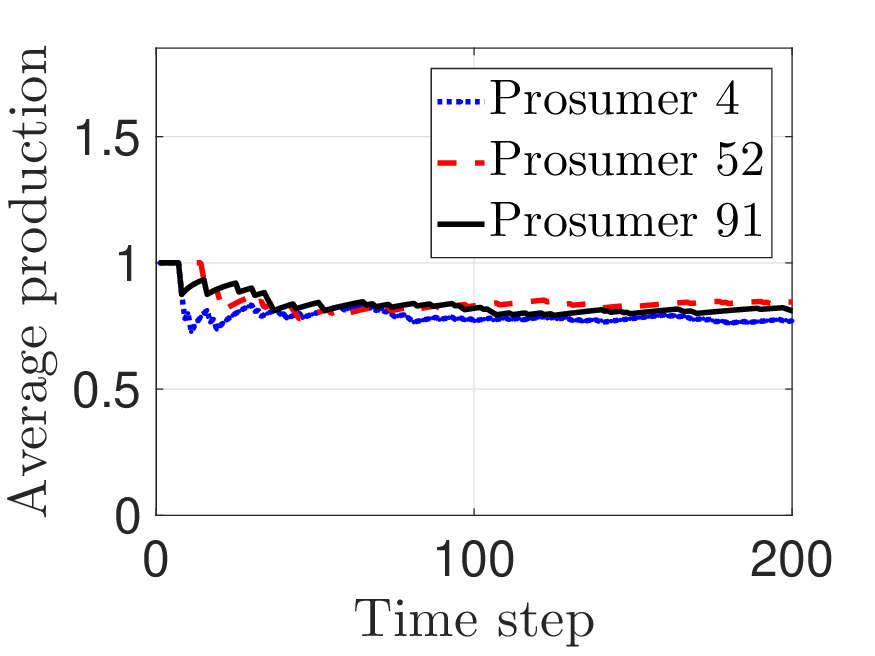}}
	\hfill
	\caption{(a) Evolution of time-averaged consumption of the resource, and (b) evolution of time-averaged production of the resource.}
	\label{fig1} 
\end{figure}
\begin{figure}[h] 
	\centering
	\subfloat[]{%
		\includegraphics[width=0.485\linewidth]{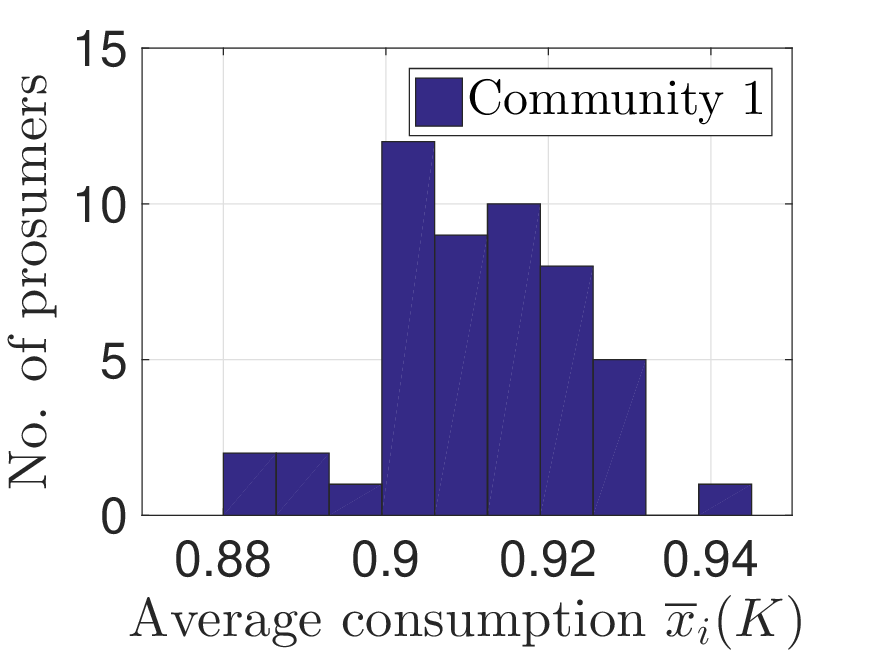}}
	\hfill
	\subfloat[]{%
		\includegraphics[width=0.485\linewidth]{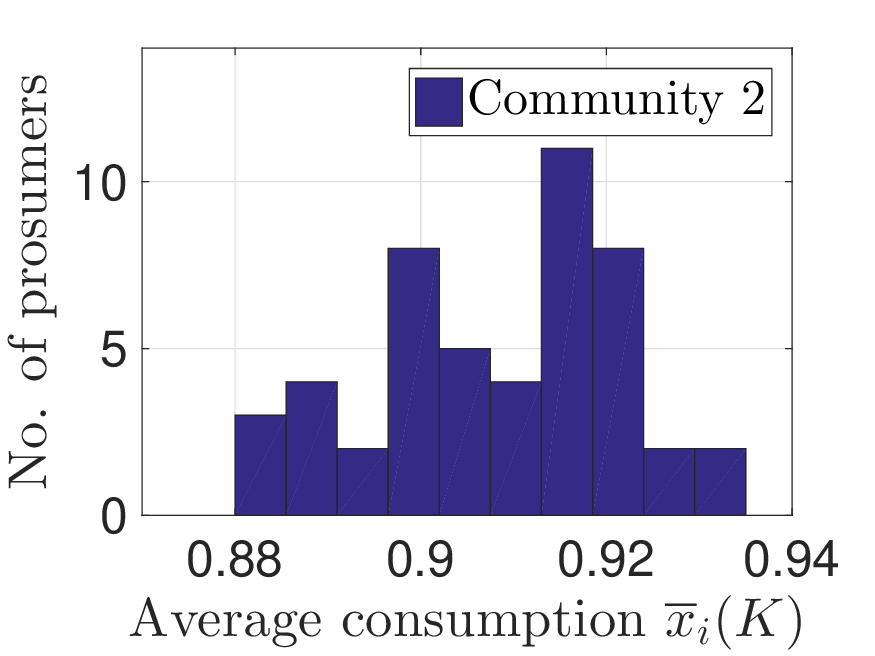}}
	\hfill
	\subfloat[]{
		\includegraphics[width=0.485\linewidth]{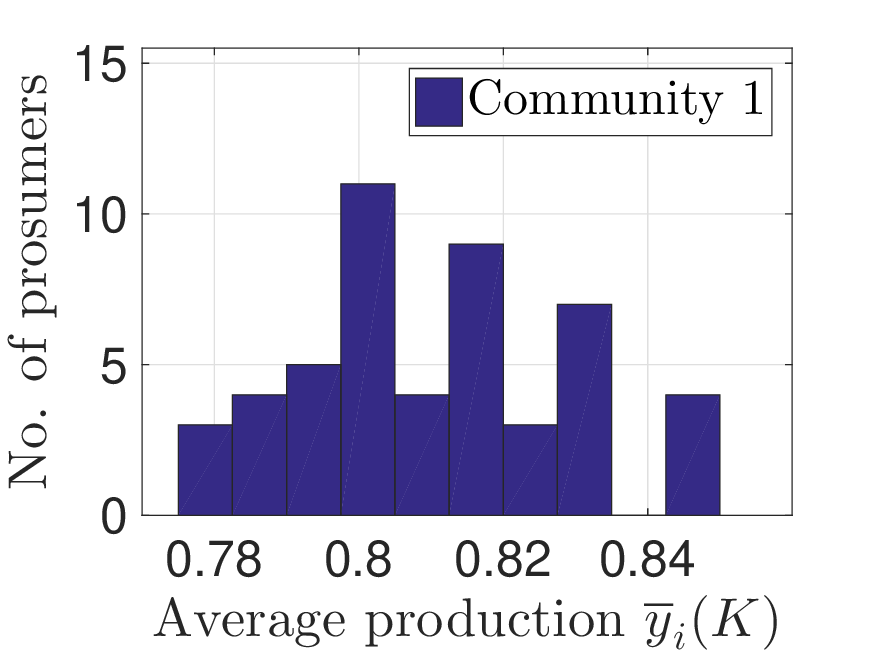}}
	\hfill
	\subfloat[]{
		\includegraphics[width=0.485\linewidth]{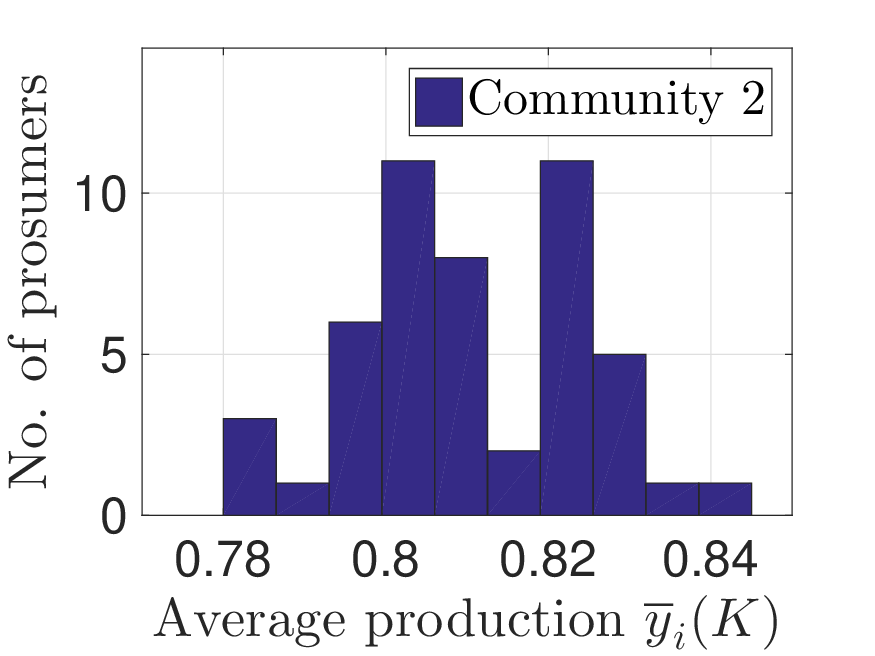}}
	\hfill
	\caption{At time index $K=200$---(a) average consumption $\overline{x}_i(K)$ by prosumers of Community $1$, (b) average consumption $\overline{x}_i(K)$ by prosumers of Community $2$, (c) average production $\overline{y}_i(K)$ by prosumers of Community $1$, and (d) average production $\overline{y}_i(K)$ by prosumers of Community $2$.}
	\label{fig1_1} 
\end{figure}
\begin{figure}[h] 
	\centering
	\subfloat[]{%
		\includegraphics[width=0.52\linewidth]{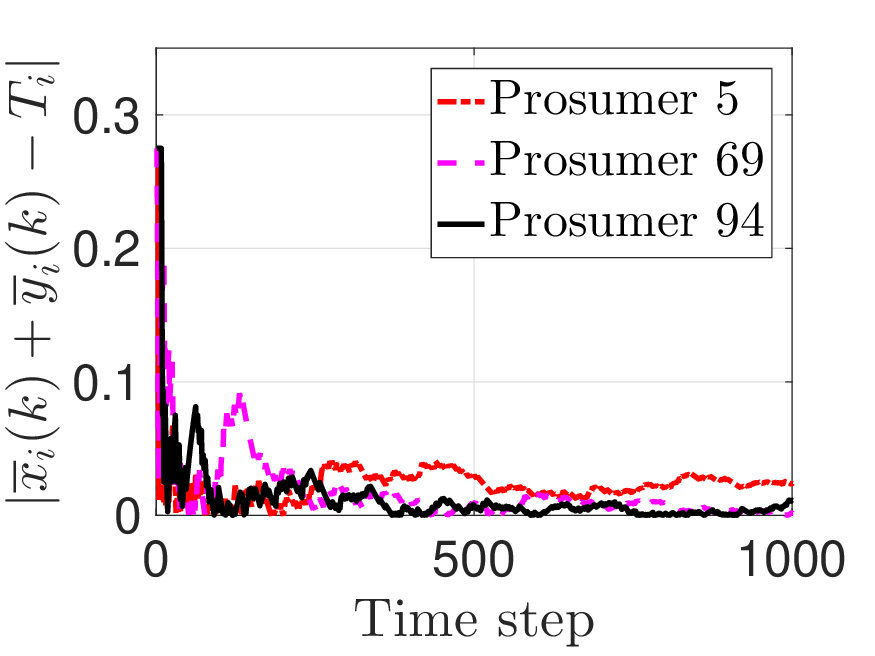}}
	\hfill
	\subfloat[]{%
		\includegraphics[width=0.495\linewidth]{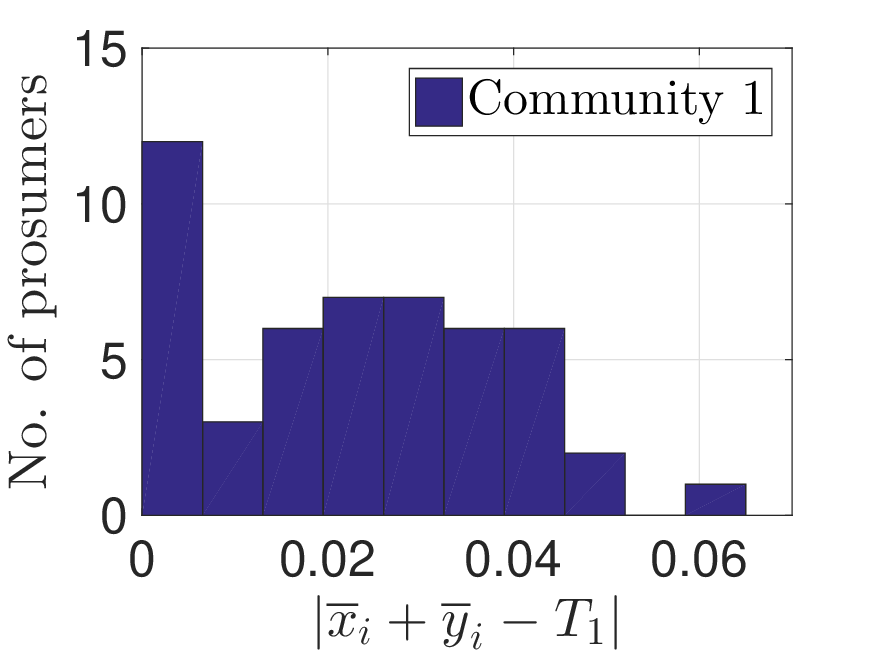}}
	\hfill
	\subfloat[]{
		\includegraphics[width=0.495\linewidth]{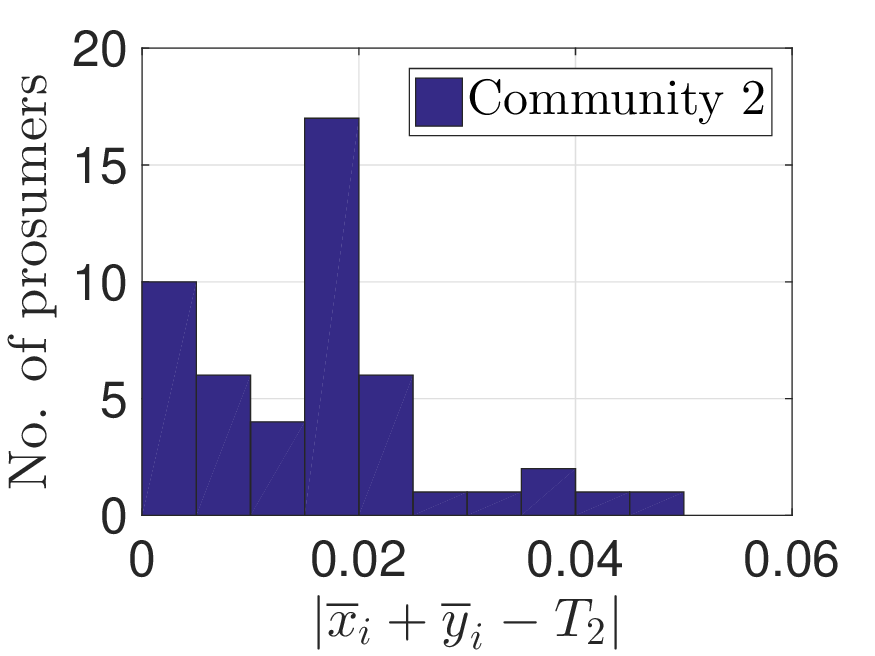}}
	\hfill
	\caption{(a) Evolution of absolute difference between desired value of utilization $T_i$ and actual utilization of the resource $\overline{x}_i(k)+\overline{y}_i(k)$ of individual prosumers, (b) absolute difference between desired value of utilization and actual utilization $|\overline{x}_i(K) + \overline{y}_i(K)- T_1|$ of prosumers of Community $1$, here $T_1 = 1.74$, and (c) absolute difference between desired value of utilization and actual utilization $|\overline{x}_i(K) + \overline{y}_i(K)- T_2|$ of prosumers of Community $2$, here $T_2 = 1.725$ and time index $K=200$.}
	\label{fig2} 
\end{figure}

\begin{figure}[h] 
	\centering
	\subfloat[]{%
		\includegraphics[width=0.495\linewidth]{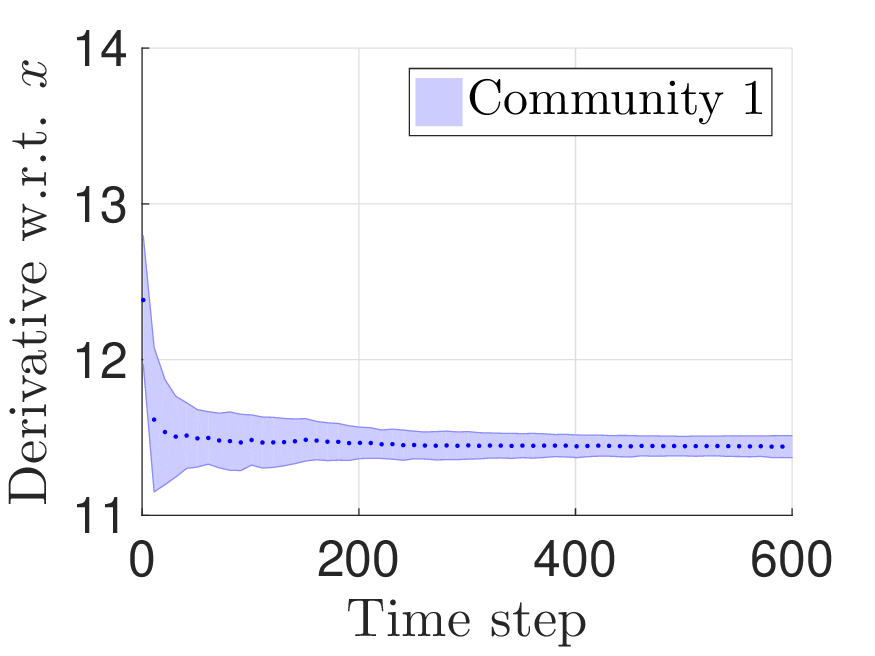}}
	\hfill
	\subfloat[]{%
		\includegraphics[width=0.495\linewidth]{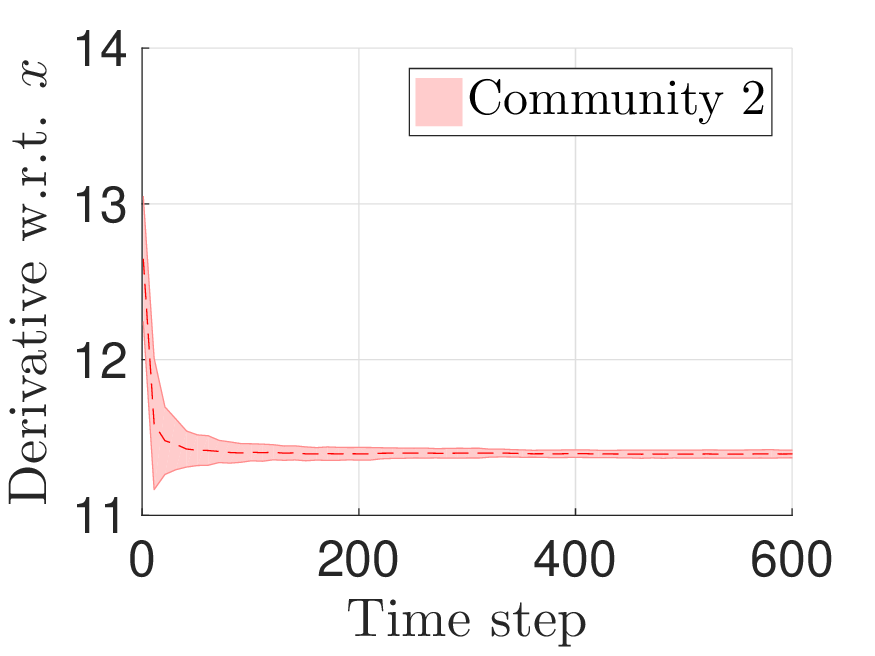}}
	\hfill
	\caption{(a) Evolution of derivatives of $g_i(.)$ w. r. t. $x$ for prosumers of Community $1$, and (b) evolution of derivatives of $g_i(.)$ w. r. t. $x$ for prosumers of Community $2$.}
	\label{fig3} 
\end{figure}
\begin{figure}[h] 
	\centering
	\subfloat[]{%
		\includegraphics[width=0.485\linewidth]{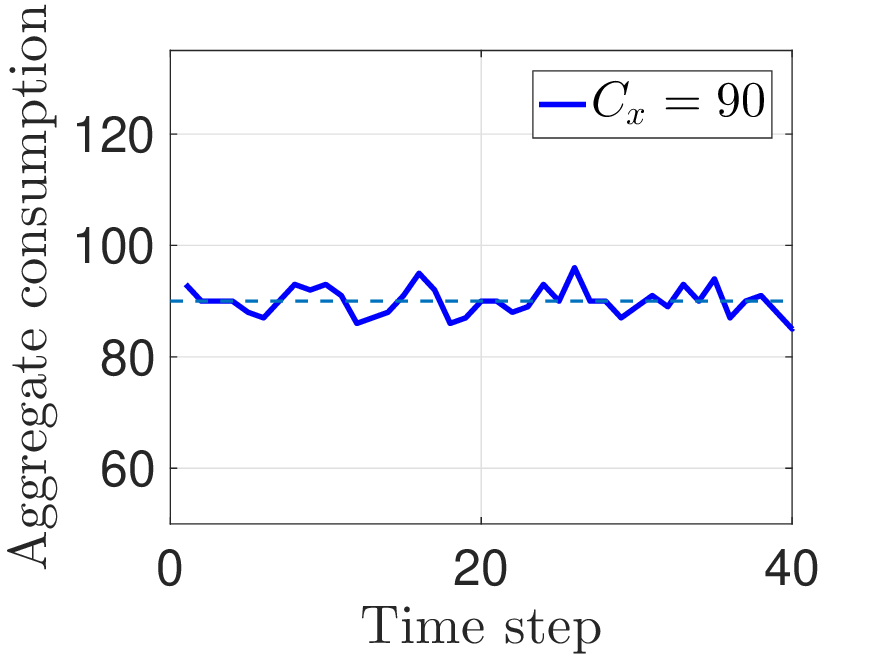}}
	\label{variance_der}\hfill
	\subfloat[]{%
		\includegraphics[width=0.485\linewidth]{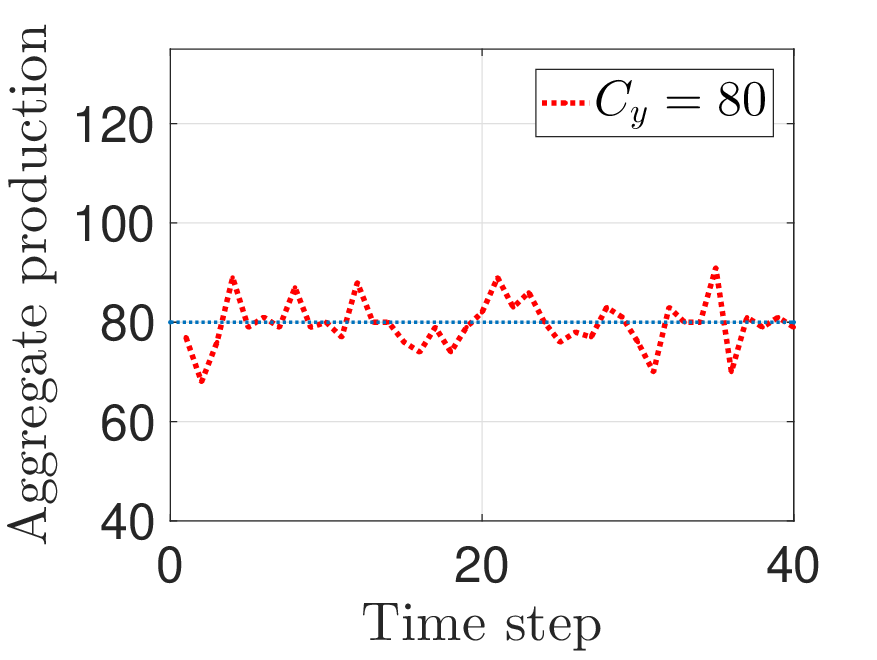}}
	\label{sum_alloc}\hfill
	\subfloat[]{%
		\includegraphics[width=0.485\linewidth]{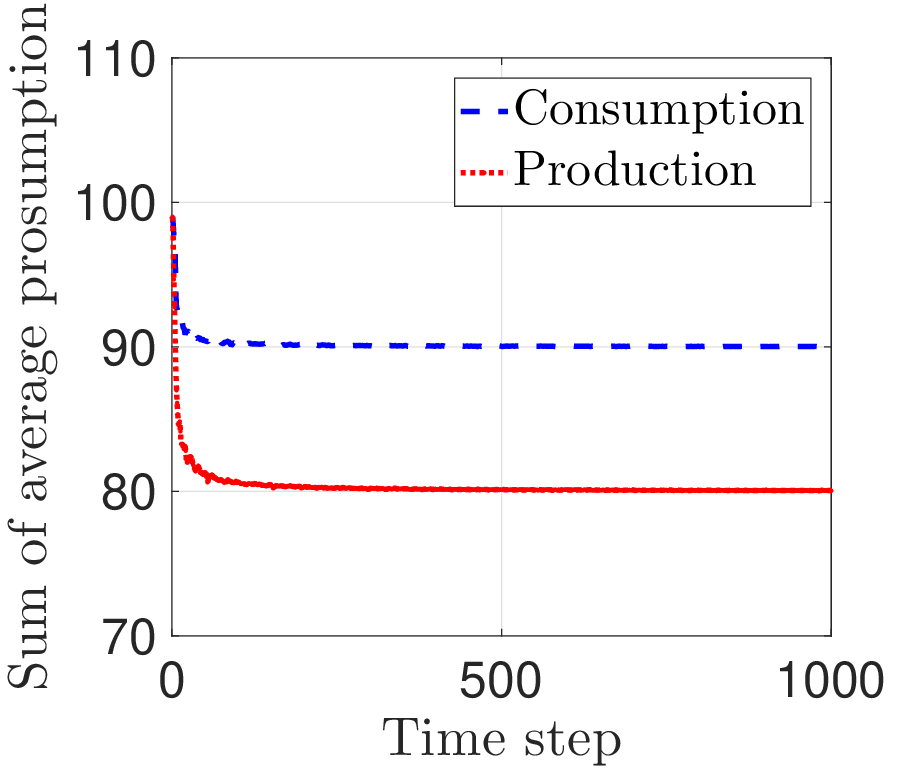}}
	\label{sum_avg_alloc}\hfill
	
	\caption{Aggregate prosumption for last $40$ time instants---(a) aggregate consumption $\sum_{i=1}^{N} x_i(k)$, (b) aggregate production $\sum_{i=1}^{N} y_i(k)$, and (c) evolution of sum of time-averaged prosumption.}
	\label{fig4} 
\end{figure}

\begin{figure}[ht]
	\centering
	
	\subfloat[]{%
		\includegraphics[width=0.485\linewidth]{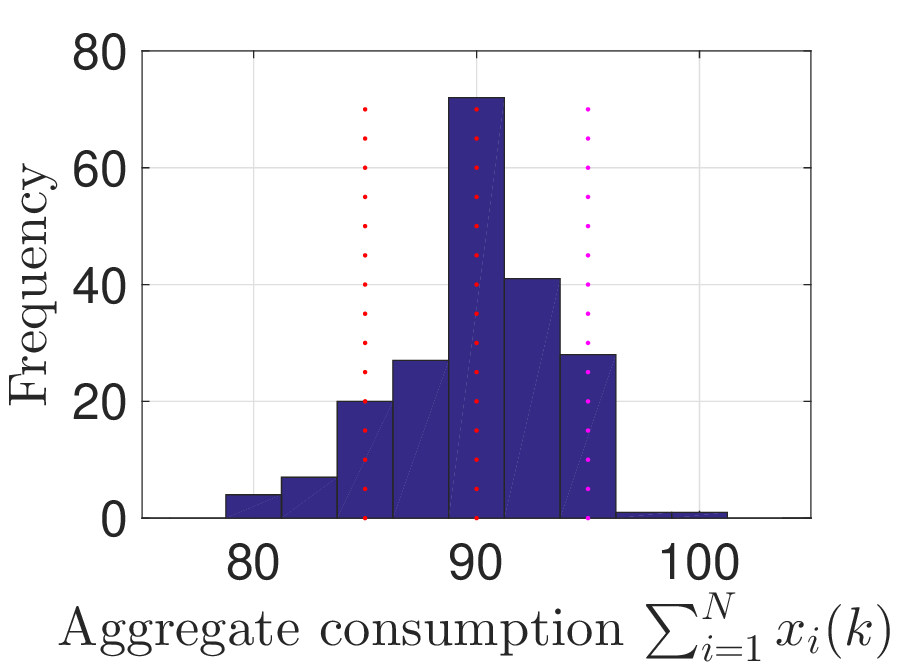}}
	\hfill
	\subfloat[]{%
		\includegraphics[width=0.485\linewidth]{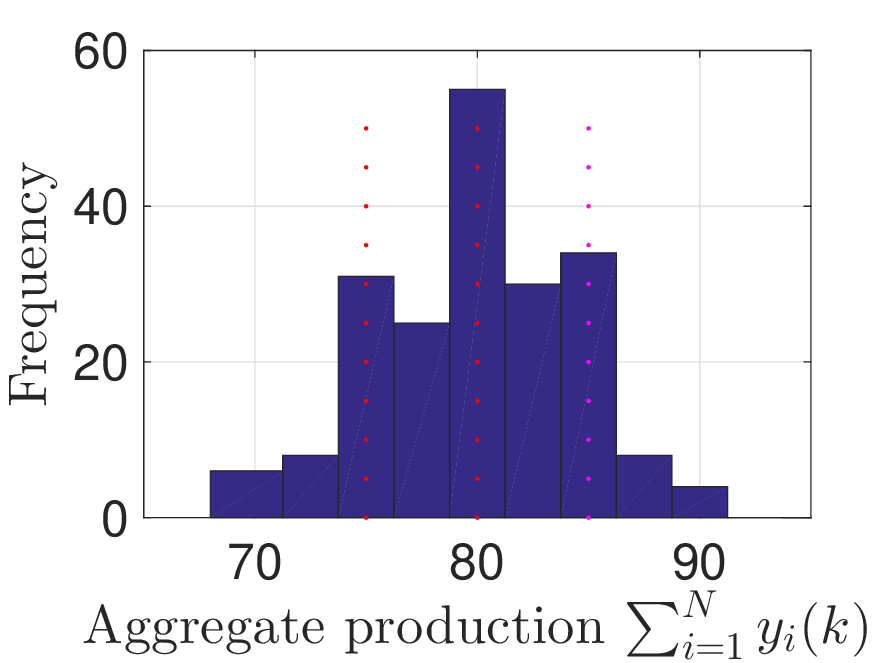}}
	\hfill	
	\subfloat[]{%
		\includegraphics[width=0.485\linewidth]{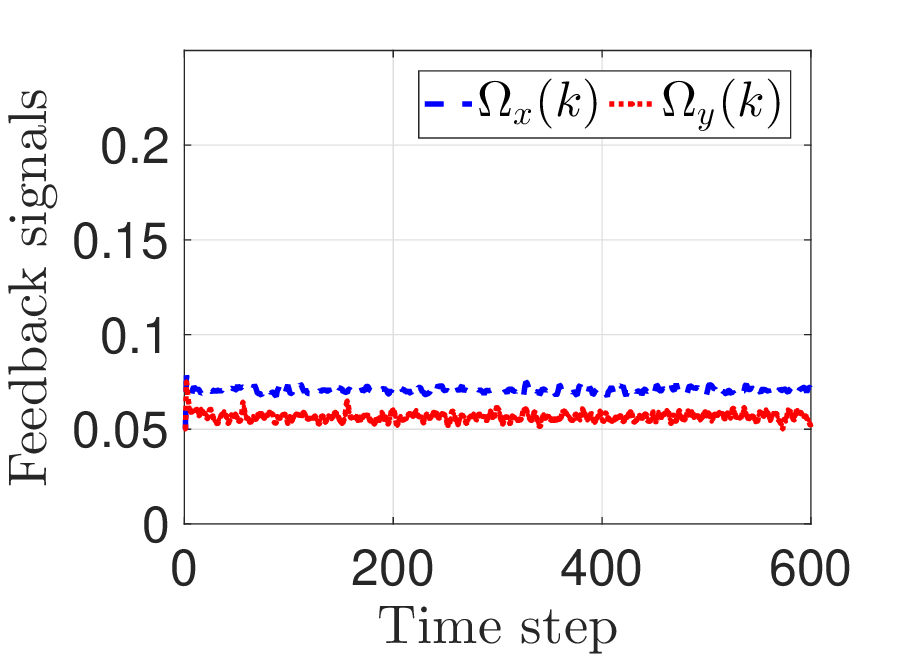}}
	\hfill
	\caption{Frequency of prosumption---(a) frequency of aggregate consumption, (b) frequency of aggregate production for last $200$ time instants, and (c) evolution of feedback signals $\Omega_x(k)$ and $\Omega_y(k)$.}
	\label{fig5} 
\end{figure}
Now, we present the simulation results and show the convergence of the long-term average of prosumption by each prosumer in Figure \ref{fig1}. Figure \ref{fig1}(a) shows the time-averaged consumption and Figure \ref{fig1}(b) shows the time-averaged production over $200$ days. In the context of car sharing, these are the average of cars used by a prosumer and the average of cars shared with another person by the same prosumer over $200$ days, respectively. Figure \ref{fig1_1} illustrates the average prosumption on the $\mathrm{200^{th}}$ day by every prosumer of a particular community. Furthermore, the absolute difference between the desired value of utilization $T_i$ of cars and the actual utilization $\overline{x}_i(k)+\overline{y}_i(k)$ by prosumer $i$ for a certain period is shown in Figure \ref{fig2}. Figure \ref{fig2}(a) illustrates the evolution of the absolute difference between $T_i$ and $\overline{x}_i(k)+\overline{y}_i(k)$ for individual prosumers. Here, we observe that gradually the difference comes closer to zero. Additionally, in Figure \ref{fig2}(b) and \ref{fig2}(c), we observe that the absolute difference between the quantities is close to zero for most of the prosumers.

Now, we analyze the derivatives of the cost functions and see, whether they gather close to each other to make consensus over time or not. Recall that the derivative of the cost function with respect to consumption is $\nabla_x g_i(.)$ and with respect to production is $\nabla_y g_i(.)$, that are shown in Figure \ref{fig3} for a single simulation.
 We plot the shaded errorbars as depicted in Figure \ref{fig3}(a) and \ref{fig3}(b). It is observed that in both the cases the derivatives gather close to each other over time. Therefore, we say that the derivatives make consensus asymptotically in their respective prosumer communities, which is a necessary and sufficient condition for optimality as described in Subsection \ref{opt_cond}. Notice that because both the derivatives (with respect to consumption and production) are same, therefore, we illustrate here just one of them. We clarify here that because of the chosen initial values, the probability $\sigma_{i,x}(k)$ may overshoot at the start of the algorithm; to keep it in the valid range, we use $\min \big \{1, \Omega_x(k) \frac{\nabla_x{g_i \big (\overline{x}_i(k),\overline{y}_i(k) \big) }}{ \overline{x}_i(k)}\big \}$. Similar step is used to keep $\sigma_{i,y}(k)$ in the valid probability range.
\newline

Now, we analyze the aggregate consumption $\sum_{i=1}^{N} x_i(k)$ and production $\sum_{i=1}^{N} y_i(k)$ by the prosumer communities. The aggregate consumption $\sum_{i=1}^{N} x_i(k)$ is presented in Figure \ref{fig4}(a) for last $40$ time instants  (days), and similarly, the aggregate production by the communities $\sum_{i=1}^{N} y_i(k)$ is shown in Figure \ref{fig4}(b). Notice that the aggregate prosumption is close to the respective capacity constraints $C_x$ and $C_y$;  overshoots and undershoots are due to the assumption of soft constraints, as described previously. Additionally, Figure \ref{fig4}(c) shows the time-averaged consumption $\sum_{i=1}^{N} \overline{x}_i(k)$ by all prosumers in the prosumer market until $1000$ time instants, and similarly, the time-averaged production $\sum_{i=1}^{N} \overline{y}_i(k)$ by all prosumers in the prosumer market for the same period, these averages are approximately equal to the respective capacities, satisfying the capacity constraints of Problem \ref{prob_des}. 
In addition to the above results, we observe in Figure \ref{fig5}(a) and \ref{fig5}(b) that most of the time the aggregate consumption $\sum_{i=1}^{N} x_i(k)$ and the aggregate production $\sum_{i=1}^{N} y_i(k)$ are close to their respective capacities $C_x$ and $C_y$. Furthermore, the convergence of feedback signals $\Omega_x$ and $\Omega_y$ are shown in Figure \ref{fig5}(c). 
\section{Conclusion and Future Directions}
We proposed distributed control algorithms to solve regulation problems with optimality constraints for community-based prosumer market. The algorithm is based on ideas from stochastic approximation but formulated in a control theoretic setting. The algorithm reaches optimality asymptotically, while simultaneously regulating instantaneous contract constraints. To do so, the algorithm does not require communication between prosumers, but little communication with the sharing platform. Additionally, the algorithm is light and is suitable to implement in an Internet-of-Things (IoT) context with minimal demands on infrastructure. Two applications are described and simulation results presented to demonstrate the efficacy of the algorithms. Future work will explore the theoretical aspects of the algorithm (convergence properties), new applications and use cases, and the development of policies to reach more complicated equilibria.

\end{document}